\pdfoutput=1

\documentclass[showpacs,preprintnumbers,amsmath,amssymb,pra,twocolumn,superscript]{revtex4}
\usepackage[utf8]{inputenc}
\usepackage{amsopn}
\usepackage{amsmath}
\usepackage{amssymb}
\usepackage{graphicx}
\usepackage{hyperref}
\newcommand{\ket}[1]{\ensuremath{|#1\rangle}}
\newcommand{\Tr}{\mathrm{Tr}}
\newcommand{\1}{{\rm 1\hspace{-0.9mm}l}}
\newcommand{\Id}{\1}

\newcommand{\dd}{\mathrm{d}}
\begin{document}
\title{Quantum control robust with coupling with an external environment}

\author{{\L}ukasz Pawela}
\email{lpawela@iitis.pl}
\affiliation{Institute of Theoretical and Applied Informatics, Polish Academy
of Sciences, Ba{\l}tycka 5, 44-100 Gliwice, Poland}

\author{Zbigniew Pucha{\l}a}
\email{z.puchala@iitis.pl}
\affiliation{Institute of Theoretical and Applied Informatics, Polish Academy
of Sciences, Ba{\l}tycka 5, 44-100 Gliwice, Poland}
\affiliation{Institute of Physics, Jagiellonian University, Reymonta 4,
30-059 Krak{\'o}w, Poland}

\begin{abstract}
We study coherent quantum control strategy which is robust with respect to
coupling with an external environment. We model this interaction by appending
an additional subsystem to the initial system and we choose the strength of the
coupling to be proportional to the magnitude of the control pulses. Therefore,
to minimize the interaction  we impose $L_1$ norm restrictions on the control
pulses. In order to efficiently solve this optimization problem we employ the
BFGS algorithm. We  use three different functions as the derivative of the $L1$
norm of control  pulses: the signum function, a fractional derivative
$\frac{\dd^\alpha  |x|}{\dd x^\alpha}$, where $0<\alpha<1$, and the Fermi-Dirac
distribution. We show that our method allows to efficiently obtain the control
pulses which neglect the coupling with an external environment.
\end{abstract}

\date{28--06--2013}



\maketitle

\section{Introduction}

The ability to manipulate the dynamics of a given complex quantum system is one
of the fundamental issues of the quantum information science. It has been an
implicit goal in many fields of science such as quantum physics, chemistry or
implementations of quantum information
processing~\cite{d2008introduction,albertini2002lie,werschnik2007quantum}. The
usage of experimentally controllable quantum systems to perform computational
task is~a very promising perspective. Such usage is possible only if a system
is controllable. Thus, the controllability of a given quantum system is an
important issue of the quantum information science, since it concerns whether it
is possible to drive a quantum system into a previously fixed state.

When manipulating quantum systems, a coherent control strategy is a widely used
method. In this case the application of semi-classical potentials, in a fashion
that preserves quantum coherence, is used to manipulate quantum states. If a
given system is controllable it is interesting to obtain control sequence which
drives a system to a desired state and simultaneously minimize the value of the
disturbance caused by imperfections of practical implementation. In the
realistic implementations of quantum control systems, there can be various
factors which disturb the evolution. One of the main issues in this context is
\emph{decoherence} -- the fact, that the systems are very sensitive to the
presence of the environment, which often destroys the main feature of the
quantum dynamics. Other disturbance can be a result of the restriction on the
frequency spectrum of acceptable control parameters~\cite{pawela2012quantum}.
In the case of such systems, it is not accurate to apply piecewise-constant
controls. In an experimental set up which utilizes an external magnetic
field~\cite{chaudhury2007quantum} such restrictions come into play and can not
be neglected.

In many situations, the interaction with the control fields causes an
undesirable coupling with the environment, which can lead to a destruction of
the interesting features of the system. In such situations, it is reasonable to
seek a control field with minimal total influence on a system. Depending on a
type of interaction with an environment the influence differs. In this article
we consider an interaction which is proportional to the magnitude of a control
field. To minimize the influence of an environment in such case, when the
control field, performs the desired evolution, the $L_1$ norm should be
minimized.


The paper is organized as follows. In Section~\ref{sec:model} we introduce the
model used for simulations. Section~\ref{sec:setup} describes the simulation
setup. In Section~\ref{sec:results} we show results of numerical simulations
and in Section~\ref{sec:conlusions} we draw the final conclusions.

\section{Our model}\label{sec:model}
To demonstrate a method of obtaining piecewise-constant controls, which have
minimal energy, we will consider an isotropic Heisenberg spin-$1/2$ chain of
a finite length $N$. The control will be performed on the first spin only. The
total Hamiltonian of the aforementioned quantum control system is given by
	\begin{equation}
		H(t) = H_0 + H_c(t),
	\end{equation} 
where 
	\begin{equation}
		H_0 = J \sum_{i = 1}^{N - 1} 
		(S_x^iS_x^{i+1}+S_y^iS_y^{i+1}+S_z^iS_z^{i+1}),
	\end{equation}
is a drift part given by the Heisenberg Hamiltonian. The control is performed
only on a first spin and is Zeeman-like, i.e.
	\begin{equation}
		H_c(t) = h_x(t)S_x^1 + h_y(t)S_y^1.
	\end{equation}
In the above $S_k^i$ denotes $k^{\text{th}}$ Pauli matrix which acts on the spin 
$i$.
Time dependent control parameters $h_x(t)$ and $h_y(t)$ are chosen to be
piecewise constant. Furthermore, as opposed to~\cite{heule2010local}, we do not
restrict the control fields to be alternating with $x$ and $y$, i.e. they can
be applied simultaneously (see e.g.~\cite{khaneja_optimal_2005} for similar
approach). For notational convenience, we set $\hbar = 1$ and after this
rescaling frequencies and control-field amplitudes can be expressed in units of
the coupling strength $J$, and on the other hand all times can be expressed in
units of $1/J$~\cite{heule2010local}.

The system described above is operator controllable, as it was shown
in~\cite{burgarth2009local} and follows from a controllability condition using
a graph infection property introduced in the same article. The controllability
of the described system can be also deduced from a more general condition
utilizing the notion of hypergraphs~\cite{puchala2012local}.

Since the interest here is focused on operator control sequence, a quality of
a control will be measured with the use of gate fidelity,
\begin{equation}
	F = \frac{1}{2^N} |\Tr( U_T^\dagger U(h) )|,
\end{equation}
where $U_T$ is the target quantum operation and $U(h)$ is an operation 
achieved by control parameters $h$. We choose gate fidelity as it neglects 
global phases.

In the case of disturbed system, we will measure the quality of the control by
a trace distance between Choi-Jamio\l{}kowski states, which gives an estimation
of a diamond norm.

We introduce an additional constrain on the control pulses, namely we wish to 
minimize the $L_1$ norm of control pulses
\begin{equation}
	||h_k||_1 = \sum_{i=1}^n|h_k^i|,\label{eq:norm}
\end{equation}
where $k \in \{x,y\}$ and $n$ is the total number of control pulses. In order
to make this quantity comparable with fidelity, we impose bounds on the
maximal amplitude of the control pulses. To accommodate this, we introduce the
following penalty
\begin{equation}
P = \frac{\sum_{i=1}^n |h_k^i|}{nb},
\end{equation}
where $b$ is the bound on the control pulse amplitude.
This leads to the following functional we wish to minimize
\begin{equation}
	G = (1 - \mu) P  - \mu F,\label{eq:functional}
\end{equation}
where $\mu$ is a weight assigned to fidelity.

To optimize the control pulses, we utilize the BFGS 
algorithm~\cite{press1992numerical}. In order to use this method effectively,
we need to calculate the explicit form of derivatives of Eq.~\eqref{eq:norm}. We
propose the following functions to be used as the derivative of the absolute 
value:
\begin{itemize}
	\item The \emph{signum} function:
	\begin{equation}
		\frac{\dd |x|}{\dd x} = \mathrm{sgn}(x).
	\end{equation}
	\item A fractional derivative:
	\begin{equation}
		\frac{\dd^\alpha |x|}{\dd x^\alpha} = \pm 
		\frac{\Gamma(2)}{\Gamma(2-\alpha)}x^{1-\alpha},
	\end{equation}
	where $\Gamma(x)=(x-1)!$ and we set $\alpha=0.99$.
	\item The Fermi-Dirac distribution
	\begin{equation}
		\frac{\dd |x|}{\dd x} \approx 2 \left( \frac{-1}{\exp(\frac{x}{kT}) + 
		}  + 
		0.5 \right),
	\end{equation}
	where we set $kT = 0.01$.
\end{itemize}

The signum function is the natural conclusion when one thinks about the
derivative of the  $L_1$ norm as it penalizes any non zero control pulses in
the control scheme.  To further out studies, we introduce two approximations of
the derivative of  the $L_1$ norm. The first one utilizes the idea of
fractional  derivatives~\cite{miller1993introduction}. This allows us to
achieve a  continuous function,  which quickly increases from 0 to 1 for
positive values of the argument and  decreases from 0 to -1 for negative
values. Although continuous, the function  has the drawback that control pulses
with lower magnitude are less penalized.  The penalty can be adjusted by using
the parameter $\alpha$

\begin{figure}[!h]
\centering\includegraphics{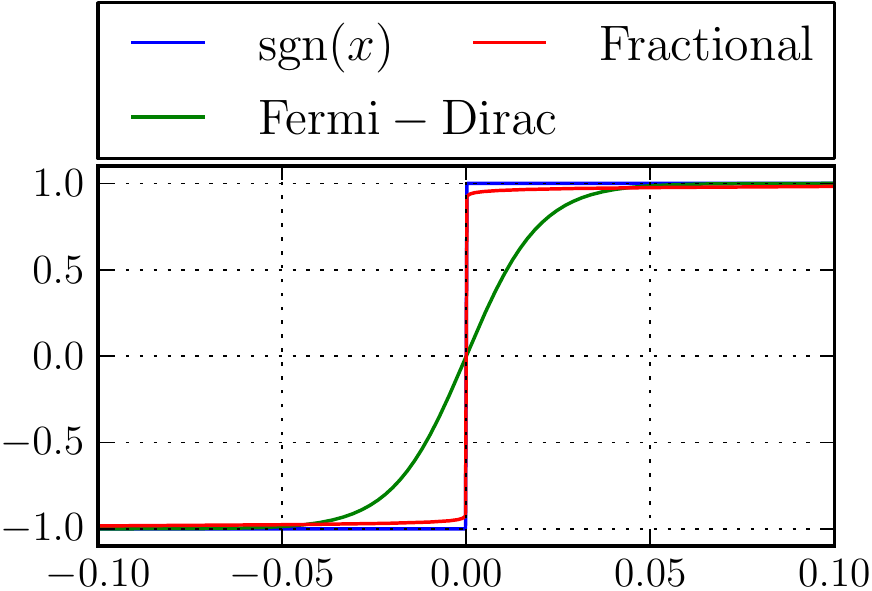}
\caption{Comparison of different derivative approximations.} 
\label{fig:derivcomp}
\end{figure}

The last proposed approximation is the Fermi-Dirac 
distribution~\cite{smirnov2007fermi}. The usage is justified, as for $T=0$ the
function is given as
\begin{equation}
f(E) = \left\{
\begin{array}{rcl}
1 & \textrm{if} & E<E_f,\\
0 & \textrm{if} & E>E_f,
\end{array}
\right.
\end{equation}
where $E_f$ is the Fermi energy. From our point of view, the function has
properties similar to the fractional derivative and the penalty for low
magnitude pulses can be adjusted by using the ``temperature'' $T$. A comparison
of these approximations is shown in Figure~\ref{fig:derivcomp}.

\section{Simulation setup}\label{sec:setup}
To demonstrate the beneficialness of our approach, we study three- and
four-qubit spin chains.
The control field is applied to the first qubit only. Our target gates are:
	\begin{equation}
		\textrm{NOT}_N = \Id^{\otimes N-1}\otimes \sigma_x,
	\end{equation}
the negation of the last qubit of the chain, and
	\begin{equation}
		\textrm{SWAP}_N = \Id^{\otimes N-2}\otimes \textrm{SWAP},
	\end{equation}
swapping the states between the last two qubits.

We provide an explicit example in which we set the duration of the control
pulse to $\Delta t = 0.2$ and the total number of pulses in each direction to
$n=64$ for the three-qubit chain and $n=256$ in the four-qubit case, although
the presented method may be applied for arbitrary values of $\Delta t$ and $n$.
The weight of fidelity in  equation~\eqref{eq:functional} is set to $\mu=0.2$
in the three qubit  scenario  and to $\mu=0.4$ in the four qubit scenario.

\section{Results}\label{sec:results}
\begin{figure}[!h]
\centering\includegraphics{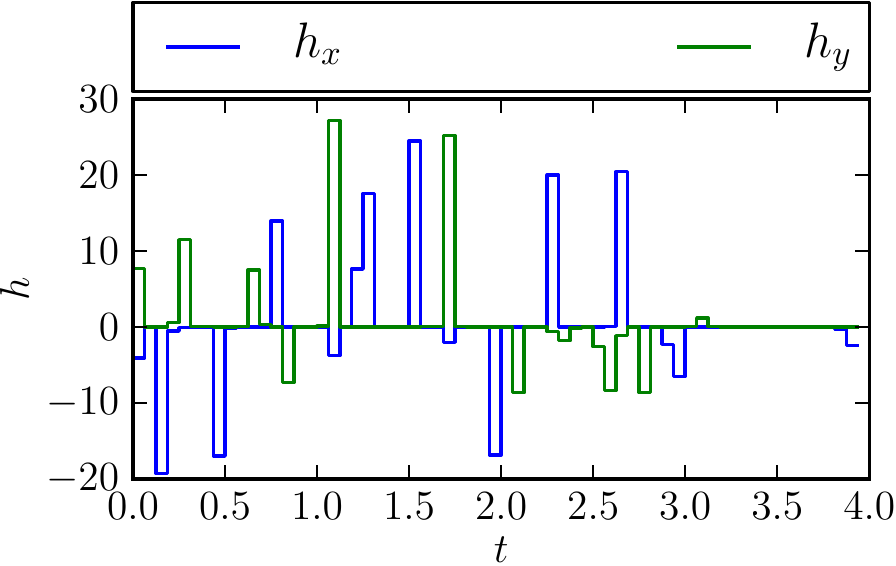}
\caption{Example control sequences $h_x$ and $h_y$ for the NOT gate in the
three qubit scenario.}\label{fig:not}
\end{figure}

\begin{figure}[!h]
\centering\includegraphics{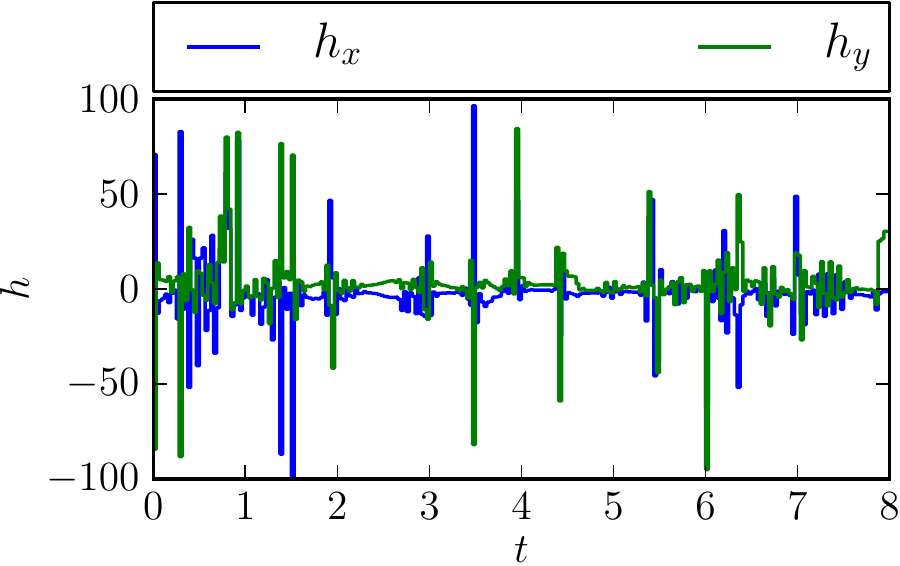}
\caption{Example control sequences $h_x$ and $h_y$ for the SWAP gate in the
four qubit scenario.}\label{fig:swap}
\end{figure}

We show examples of control sequences obtained by using our method in
Figs.~\ref{fig:not} and \ref{fig:swap}. They depict results obtained for the
three qubit NOT gate optimization and four qubit SWAP gate optimization
respectively. In the three qubit scenario we find, as expected, a control
sequence which equal to zero most of the time with irregular, high amplitude
pulses. A similar case can be made for the swap gate in the four qubit
scenario. The main difference is that in this case the high amplitude pulses
are surrounded by groups of weaker pulses. The results shown here are for the
fractional derivative approximation. Simulations for other approximation yield
nearly identical results.

The fidelity obtained in both cases is $F > 0.99$ and the value of $P$ has the
order of $10^{-2}$.

\begin{figure}[!h]
\includegraphics{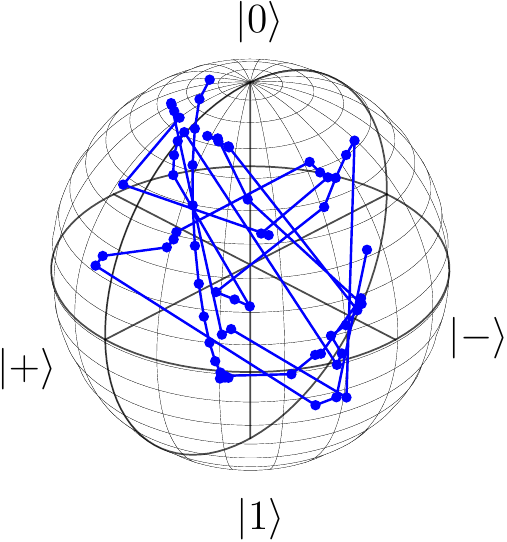}
\caption{Time evolution of the first qubit of a three qubit chain from the
state $\ket{000}$ to  the state $\ket{001}$ under the operator  $\1 \otimes \1
\otimes \sigma_x$ implemented by optimized control 
sequences.}\label{fig:evo3_1}
\end{figure}

\begin{figure}[!h]
\includegraphics{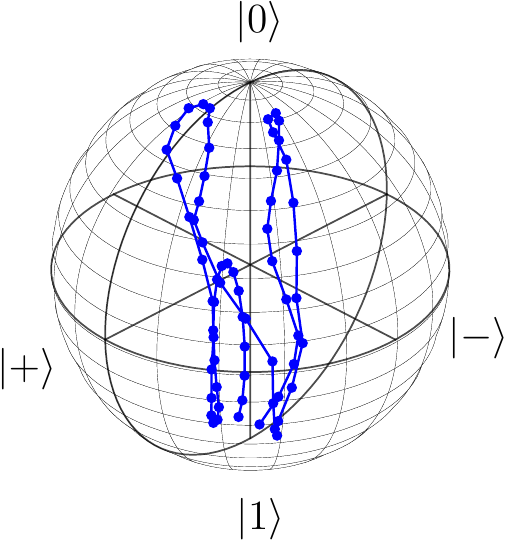}
\caption{Time evolution of the second qubit of a three qubit chain from the
state $\ket{000}$ to  the state $\ket{001}$ under the operator  $\1 \otimes \1
\otimes \sigma_x$ implemented by optimized control 
sequences.}\label{fig:evo3_2}
\end{figure}

\begin{figure}[!h]
\includegraphics{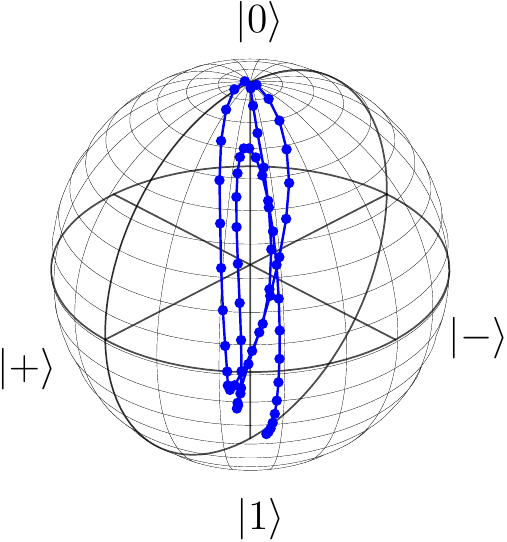}
\caption{Time evolution of the third qubit of
a three qubit chain from the  state $\ket{000}$ to  the state $\ket{001}$ under
the operator  $\1 \otimes \1 \otimes \sigma_x$ implemented by optimized control
sequences.}\label{fig:evo3_3}
\end{figure}

\begin{figure}[!h]
\includegraphics{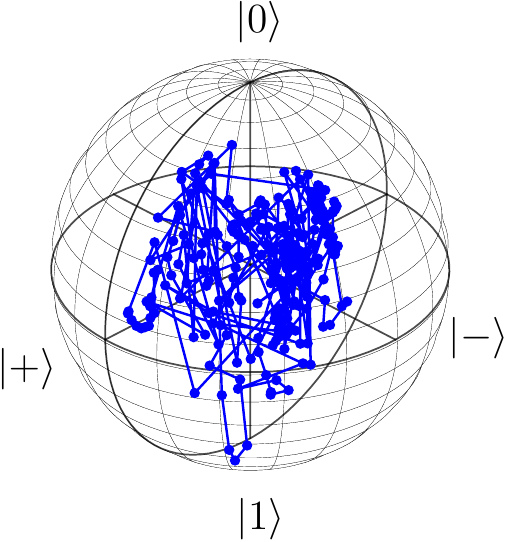}
\caption{Time evolution of the first qubit of a four qubit chain from the
state $\ket{0010}$ to the state $\ket{0001}$ under the operator
$\1 \otimes \1 \otimes \mathrm{SWAP}$ implemented by optimized control
sequences.}\label{fig:evo4_1}
\end{figure}

\begin{figure}[!h]
\includegraphics{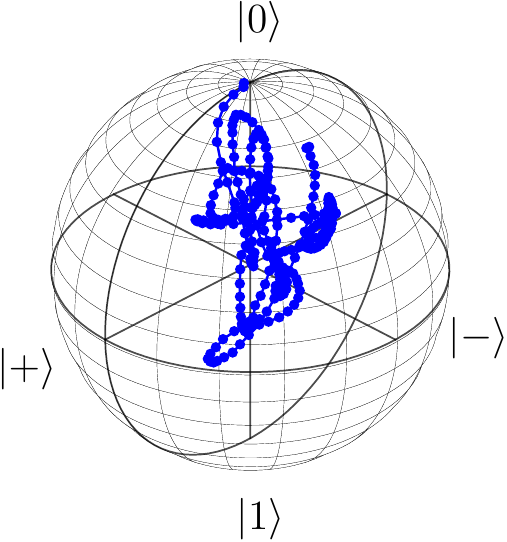}
\caption{Time evolution of the second qubit of a four qubit chain from the
state $\ket{0010}$ to the state $\ket{0001}$ under the operator
$\1 \otimes \1 \otimes \mathrm{SWAP}$ implemented by optimized control
sequences.}\label{fig:evo4_2}
\end{figure}

\begin{figure}[!h]
\includegraphics{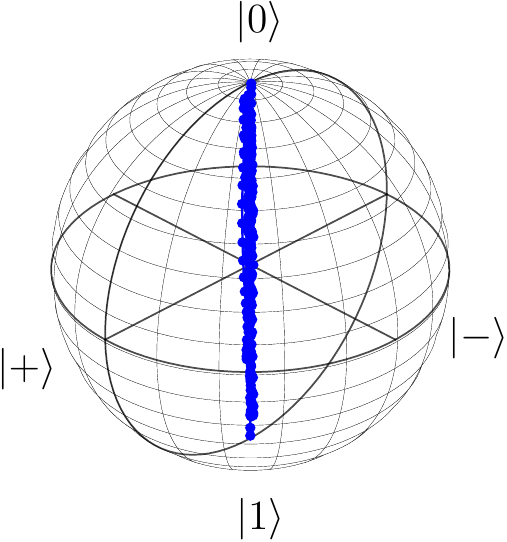}
\caption{Time evolution of the third qubit of a four qubit chain from the
state $\ket{0010}$ to the state $\ket{0001}$ under the operator
$\1 \otimes \1 \otimes \mathrm{SWAP}$ implemented by optimized control
sequences.}\label{fig:evo4_3}
\end{figure}

\begin{figure}[!h]
\includegraphics{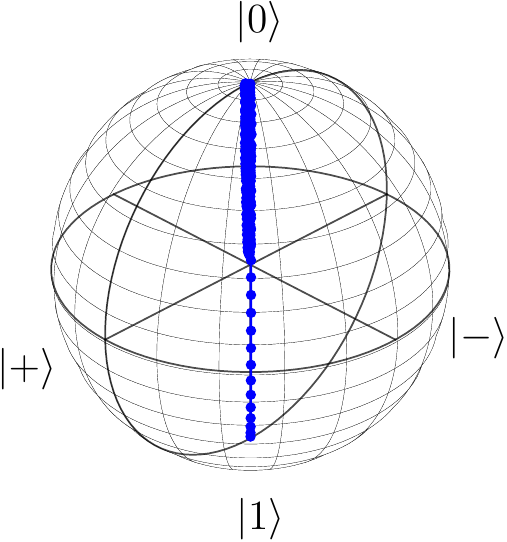}
\caption{Time evolution of the fourth qubit of a four qubit chain from the
state $\ket{0010}$ to the state $\ket{0001}$ under the operator
$\1 \otimes \1 \otimes \mathrm{SWAP}$ implemented by optimized control
sequences.}\label{fig:evo4_4}
\end{figure}

Finally, we show the evolution of each qubit's state. Let the qubits be in the
state $\ket{\psi}_0=\ket{000}$ in the case of the three qubit scenario.
Figs.~\ref{fig:evo3_1}, \ref{fig:evo3_2} and \ref{fig:evo3_3} show the time
evolution of each qubit state in this  setup. The final state of the chain is
$\psi_f=\ket{001}$. In the four qubit  scenario the time evolution is shown in
Figs.~\ref{fig:evo4_1}, \ref{fig:evo4_2}, \ref{fig:evo4_3} and
\ref{fig:evo4_4}. Let the initial state of the chain be equal to $\ket{\phi_0} 
= \ket{0010}$. The final state of the chain is $\ket{\phi} = \ket{0001}$.



In order to demonstrate the advantages of our approach, we perform additional
simulations, where we put $\mu=1$ in Eq.~\eqref{eq:functional}. This is the
unconstrained problem of finding optimal control pulses. Next, we introduce an
interaction with an environment, proportional to $|h_x| + |h_y|$. We model the
interaction with the environment by adding a qubit  to  the chain. The
Hamiltonian for this case is
\begin{equation}
\begin{split}
H_\mathrm{graph}(t) =& H_0 + H_\mathrm{c}(t)+ \gamma (|h_x| + |h_y|)\times \\
&\times \sum_{i=1}^N 
(S_x^iS_x^{N+1}+S_y^iS_y^{N+1}+S_z^iS_z^{N+1}).
\end{split}
\end{equation}
In order to compare the evolution with the additional qubit with a given $U_T$
we use the following scheme. For a quantum channel $\Phi$, let us write
$J(\Phi)$ to denote the associated state:
\begin{equation}
J(\Phi) = \frac{1}{n} \sum_{1\leq i,j \leq n} \Phi(|i \rangle \langle j|)
\otimes |i \rangle \langle j|.
\end{equation}
Here we are assuming that the channel maps $n\times 
n$ complex matrices into $m\times m$ complex matrices. The matrix $J(\Phi)$ is
sometimes called the Choi-Jamio{\l}kowski representation of $\Phi$.
For quantum channels $\Phi_0$ and $\Phi_1$ we may 
define the "diamond norm distance" between them as
\begin{equation}
\| \Phi_0 - \Phi_1 \|_{\Diamond} =
\sup_{k,\rho} \: \| (\Phi_0 \otimes \1_k)(\rho) - (\Phi_1 \otimes 
\1_k)(\rho) \|_1\label{eq:diamond}
\end{equation}
where $\1_k$ denotes the identity channel from  the set of $k \times k$ complex
matrices to itself, $\| \cdot \|_1$ denotes the trace norm, and the supremum is
taken over all $k \geq 1$ and all density matrices $\rho$ from the  set of $nk
\times nk$ complex matrices.  The supremum always happens to be achieved for
some choice of $k\leq n$ and some rank 1 density matrix $\rho$. A coarse bound
for the diamond norm defined in Eq.~\eqref{eq:diamond} is known~\cite{kitaev2002classical}
\begin{equation}
\frac{1}{n} \| \Phi_0 - \Phi_1 \|_{\Diamond} \leq
\| J(\Phi_0) - J(\Phi_1) \|_1 \leq \| \Phi_0 - \Phi_1 
\|_{\Diamond}.\label{eq:distance}
\end{equation}
Therefore, to compare the target operations with and without the  additional
qubit, we study the $L_1$ of the difference of the Jamio{\l}kowski  matrices of
the respective quantum channels $\| J(\Phi_0) - J(\Phi_1) \|_1$.  The results
for different  target  operations are summarized in Tab.~\ref{table:summary}. We
show results obtained for Fermi-Dirac approximation of the derivative. As stated
in the table, the bigger the system under consideration is the greater is the
gain from using our method.
\begin{table}[h!]
	\begin{tabular}{c|c|c||c|c|}
	\cline{2-5}
	& \multicolumn{2}{|c||}{Without additional qubit} & 
	\multicolumn{2}{|c|}{With additional qubit} \\ \cline{2-5}
	& $\mu=1$ & $\mu < 1$ & $\mu=1$ & $\mu < 1$  \\ \hline
	\multicolumn{1}{|c|}{NOT$_3$} & 0.0000 & 0.0000 & 0.0975 & 0.0086 \\ \hline
	\multicolumn{1}{|c|}{NOT$_4$} & 0.0000 & 0.004 & 0.9788 & 0.0142 \\ \hline
	\multicolumn{1}{|c|}{SWAP$_3$} & 0.0000 & 0.0001 & 0.0135 & 0.0133 \\
	\hline
	\multicolumn{1}{|c|}{SWAP$_4$} & 0.0000 & 0.0020 & 0.0843 & 0.0064 \\
	\hline
	\end{tabular}
	\caption{Summary of the value of Eq.~\eqref{eq:distance} for teh studied 
	cases. For $\mu=1$ we 
	have a control optimization without regarding the $L_1$ norm of control 
	pulses.}\label{table:summary}
\end{table}
\vspace{1.5cm}

 \section{Conclusions}\label{sec:conlusions}
 In this work we introduced a method of obtaining a piecewise constant control
field for a quantum system with an additional constrain of minimizing the $L_1$
norm. To demonstrate the beneficialness of our approach, we have shown results
obtained for a spin chain, on which we implemented two quantum operations:
negation of the last qubit of the chain and swapping the states of the two last
qubits of the chain. Our results show that it is possible to obtain control
fields which have minimal energy and still give a high fidelity of the quantum
operation. Our method may be used in situations where the interaction with the
control field causes additional coupling to the environment. As our method
allows one to minimize the number of control pulses, it also minimizes the
amount of coupling to the environment. Other possible usage of our method
includes systems, in which it is possible to use rare, but high value of control
pulses, like for example superconducting magnets with high impulse current.

\begin{acknowledgements}
{\L}.~Pawela was supported by the Polish National Science Centre under the
grant number N N514 513340.
Z.~Pucha{\l}a was supported by the Polish Ministry of Science and Higher
Education under the project number IP2011 044271.
\end{acknowledgements}

\bibliography{minimal_energy}
\bibliographystyle{apsrev}
\end{document}